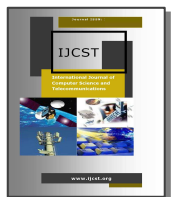

# Effect of the Feedback Function on the QoS in a Time Space Priority with Active Queue Management

Mohamed HANINI[1,3], Said EL KAFHALI[1,3], Abdelkrim HAQIQ[1,3] and Amine BERQIA[2,3]

[1]Computer, Networks, Mobility and Modeling laboratory, Department of Mathematics and Computer
FST, Hassan 1st University, Settat, Morocco
[2]Learning and Research in Mobile Age team (LeRMA), ENSIAS, Mohammed V Souissi University, Rabat, Morocco
[3]e-NGN Research group, Africa and Middle East
{haninimohamed, kafhalisaid, ahaqiq, berqia}@gmail.com

*Abstract*— The basic idea behind an active queue management (AQM) is to sense the congestion level within the network and inform the packet sources about, so that they reduce their sending rate. In literature a lot off mechanisms of AQM are studied. But there are not used in the context of the DiffServ architecture where different types of packet with different requirements of QoS share the same link. In this paper, we study an access control mechanism for RT and NRT packets arriving in a buffer implemented at an end user in HSDPA. The mechanism uses thresholds to mange access in the buffer and gives access priority to RT packets. In order to control the arrival rate of the NRT packets in the buffer an active queue management is used. We study the effect of the feedback function on the QoS parameters for both kinds of packets .Mathematical description and analytical results are given, and numerical results show that the proposed function achieves higher QoS for the NRT packets in the system.

*Index Terms*— Active Queue Management, Congestion Control HSDPA, QoS and Queueing Theory

## I. INTRODUCTION

THE phenomenal expansion of communication networks has seen a rapid growth in the number and variety of applications. Under this network development, the control of congestion has become one of the most critical issues in present networks to accommodate the increasingly diverse rang of services and types of traffic [5]. It is also a major challenge to the researchers in the field of performance modelling. Congestion control to enable different types of traffic to satisfy specified quality of service (QoS) constraints is becoming significantly important [14].

While third Generation (3G) wireless systems are being intensively deployed worldwide, new proposals for enhanced data rates and quality of service (QoS) provision are being standardized. One of the most promising enhancements to the widely deployed Universal Mobile Telecommunication System (UMTS) which is based on Wideband Code Division Multiple Access (W-CDMA) is called High Speed Downlink Packet Access (HSDPA) [1], which is also referred to as 3.5G.

HSDPA aims at providing data rates around 2-10Mbps on the downlink to mobile users mainly for multimedia services such as real-time and streaming video in packet-switched common channel.

To achieve these requirements, three key techniques are added in release 5: Adaptive Modulation and Coding (AMC), Hybrid Automatic Repeat Request (HARQ) and Fast Scheduling [1]. One important change from release 99 channels is the location of scheduler at Node B as opposed to the Radio Network Controller (RNC), thus we can implement buffer in node B to improve requirement of QoS for both RT and NRT packets [16].

The target of majority of packet scheduling algorithms proposed and studied in the literature [7, 12, 21, 19] is to distribute the time resources between the users.

However HSDPA link supports multimedia services, which require differentiated QoS. Thus, besides user priority, the algorithm also has to determine priority among classes supported for a given user. The works in [13, 27] were interested to integrate a buffer per user in node B to take into account intra-user traffic differentiation.

Authors in [24] present a modeling for a multimedia traffic in a shared channel, but they take in consideration system details rather the characteristics of the flows composing the traffic.

Works in [2], [8], [20] study priority schemes and try to maximize the QoS level for the RT packets, without taking into account the effect on degradation of the QoS for NRT packets.

Therefore, Suleiman et all present in [25] a queuing model for multimedia traffic over HSDPA channel using a combined time priority and space priority (TSP priority) with threshold to control QoS measures of the both RT and NRT packets.

The basic idea of TSP priority [26] is that, in the buffer, RT packets are given transmission priority (time priority), but the number accepted of this kind of packets is limited. Thus, TSP scheme aims to provide both delay and loss differentiation.

To model the TSP scheme, mathematical tools have been used in [27] and QoS measures have been analytically deducted, but some given results are false, [9] and [10] corrected this paper and used MMPP and BMAP processes to model the traffic sources.

An enhanced algorithm (Enhanced Basic TSP: EB-TSP) is studied in [15], the authors propose to modify the priority function to overcome the bad management for buffer space in [27] the proposed model improve QoS for RT packet by reducing the loss probability of RT packets, and by achieving a better management for the network resources.

The extension studied in [25], [26] incorporates thresholds to control the arrival packets of NRT packets (Active TSP scheme), and show, via simulation (using OPNET), that TSP scheme achieves better QoS measures for both RT and NRT packets compared to FCFS (First Come First Serve) queuing.

This extension is based on the principle of Active Queue Management (AQM). The basic idea behind an AQM algorithm is to sense the congestion level within the network and inform the packet sources about this so that they reduce their sending rate [11, 23, 18, 4].

In the opposition of the reduction adopted in [25], where the NRT arrival rate is reduced to an arbitrary value, we propose in this work a reduction which is a function of the total number of the packets in the buffer. And we show that this proposition improve the performance parameters of NRT packets.

## II. SYSTEM DESCRIPTION AND MATHEMATICAL ASSUMPTIONS

In the mechanism studied in this paper, we model the HSDPA downlink to an end user by a queue of finite capacity N. The arriving flow to the user is heterogeneous composed by RT and NRT packets.

The arrivals of RT and NRT packets are assumed to be poisson processes with rates $\lambda$ and $\lambda_1$ respectively. The service times of RT and NRT packets are assumed to be exponential with rate $\mu$ and $\mu_1$ respectively.

We also assume that the arrival processes and the service times are mutually independent between them.

The state of the system at any time t can be described by the process $X(t) = (X_1(t), X_2(t))$,

where $X_1(t)$ (respectively $X_2(t)$) is the number of RT (respectively of NRT) packets in the buffer at time t.
The state space of $X(t)$ is $E=\{0,...., R\} \times \{0,...., H\}$.

The RT packets have priority in the queue and their number cannot exceed a given threshold R (R < N), all arrived RT packets after this threshold will be lost.

Two other thresholds $L$ and $H$ (R < L < H such that H=N-R) are used in order to control the arrival rate of the NRT packets (Figure 1). The condition: $H = N - R$, permits to avoid the NRT packet loss.

The arrival rate of NRT packets is reduced in a linear manner according to $k$ (where: $k$ is the total number of packets in the queue):

- If $0 \leq k < L$, then the arrival rate of the NRT packets is $\lambda_1$.
- If $L \leq k < H$ then the arrival rate of NRT packets is $\lambda_1(k)$ which is a linear decreasing function from $\lambda_1$ when $k = L$ to 0 when $k = H$
- If $k \geq H$, then no NRT arrives in the system.

This can be considered as an implicit feedback from queue to the Node B.

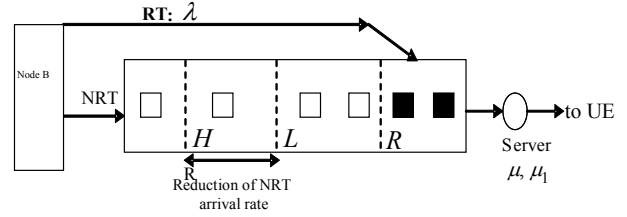

Fig. 1: Queuing model applied to end user

Between the thresholds $L$ and $H$, the reduction of the arrival rate of NRT packets from $\lambda_1$ to a lower value is defined as following (Figure 2):

$$\lambda_1(k) = \frac{H-k}{H-L}.\lambda_1 \quad \text{where:}$$
$$k \in \{i + j /(i, j) \in E \text{ and } L \leq i + j \prec H\}$$

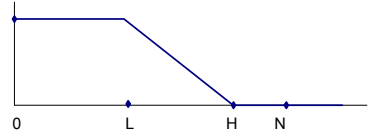

Fig. 2: Linear reduction of arrival rate of NRT packets versus total number of packets in the queue

## III. STEADY STATE AND PERFORMANCE PARAMETERS

### A. Steady State and Equilibrium Probability

For the system, the arrival processes are Poisson, therefore, the inter-arrivals are exponentials. Moreover, the servers are exponentials and all these variables are mutually independent, hence $X(t)$ is a Markov process with finite state spaces E.

$X(t)$ is also irreducible, since all states communicate. We deduct that $X(t)$ is ergodic (stable).

Consequently, the equilibrium probabilities of $X(t)$ exist. We denote the equilibrium probability of X(t) at the state (i,j) by $\{p(i,j)\}$, we know that:
$$p(i, j) = \lim_{t \to +\infty} p(X_1(t) = i, X_2(t) = j)$$

It can be computed by solving the system of the balance equations given in the appendix (the average flow outgoing of each state is equal to the average flow go into this state) and adding the normalization equation (the sum of all state probabilities equal to 1).

Note that in these equations: $\lambda_1(k) = (a.k + b).\lambda_1$ where :

$a = \dfrac{-1}{H - L}$ and $b = \dfrac{H}{H - L}$.

*B. QoS Parameters*

In this section, we determine analytically, different performance parameters (loss probability of the RT packets, average numbers of the RT and NRT packets in the queue, and average delay for the RT and NRT packets) at the steady state.

These parameters can be derived from the stationary state probabilities as follows:

  a) *Loss probability of the RT packets :*

It refers to the percentage of data loss among all the delivered data in the steady state

$$P_{LRT} = \sum_{j=0}^{H} p(R, j)$$

  b) *Average numbers of the RT packets in the queue:*

It is given by

$$N_{RT} = \sum_{i=0}^{R}\sum_{j=0}^{H} i\, p(i, j)$$

  c) *Average numbers of the NRT packets in the queue:*

Is obtained as follows:

$$N_{NRT} = \sum_{j=0}^{H}\sum_{i=0}^{R} j\, P(i, j)$$

  d) *Average Packets Delay*

It is defined as the number of packets waits in the queue since its arrival before it is transmitted. We use Little's law [22] to obtain respectively the average delays of RT and NRT packets in the system as follows:

$$D_{RT} = \dfrac{N_{RT}}{\lambda .(1 - P_{LRT})}$$

$$D_{NRT} = \dfrac{N_{RT} + N_{NRT}}{\lambda_{eff}(NRT)}$$

where the term $\lambda_{eff}(NRT)$ indicates the effective arrival rate of the NRT packets in the system, it can be obtained from

$$\lambda_1.\sum_{i=0}^{R}\sum_{j=0}^{L-i-1} p(i,j) + \lambda_1.\sum_{i=0}^{R}\sum_{j=L-i}^{H-i-1}(a(i+j)+b)p(i,j)$$

## IV. NUMERICAL RESULTS

In this section we present the numerical results for the proposed model. We compare the effect of the linear reduction of $\lambda_1$ on the performance parameters with the case where the reduction is to an arbitrary value ($\dfrac{\lambda_1}{2}$ or $\dfrac{\lambda_1}{4}$).

Numerical parameters used are summarized in Table 1.

Table1: Summary of numerical parameters

| | |
|---|---|
| $N$ | 100 |
| $L$ | 50 |
| $\lambda$ | 30 |
| $\mu$ | 30 |
| $\mu_1$ | 35 |
| $H$ | N-R |

*A. Impact of $\lambda_1$ on NRT Performance Parameters*

To show the effect of the proposed approach, we calculate the average delay for NRT packets for different reduction arbitrary value, and we use $R = 30$.

For reduction to $\dfrac{\lambda_1}{2}$, which represents the mean value of the linear reduction proposed, Figure3 show that the linear reduction permits to achieve a better average delay for NRT packets.

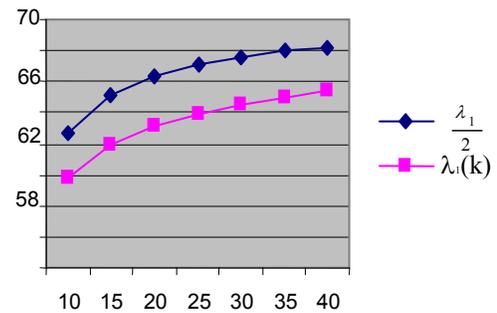

Fig. 3: Number of NRT packets versus arrival rate of NRT packets

When this arbitrary reduction between thresholds $L$ and $H$ becomes lower: $\dfrac{\lambda_1}{4}$, Figure4 show that: for the lower value of $\lambda_1$, the arbitrary reduction achieves a good delay for NRT packets in comparison with the linear reduction. But the model with linear reduction permits to achieve lower delay for NRT packets when $\lambda_1$ takes higher value.

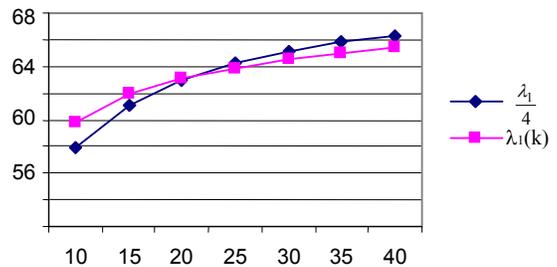

Fig. 4: Number of NRT packets versus the value of arrival rate of NRT packets

## B. *Impact of the Threshold $R$*

Varying threshold $R$ means that the space reserved to NRT packets varies. To show the impact of increasing $R$ on the average delay for NRT packets, the numerical results bellow has been done for the values of $\lambda_1 = 15$.

Fig. 5 confirms the results obtained before, and even if the space reserved for NRT packets is reduced (when $R$ increases), the model with linear reduction of arrival rate of NRT packets gives always lower average delay for NRT packets in comparison with the model with an arbitrary reduction equal to $\frac{\lambda_1}{2}$.

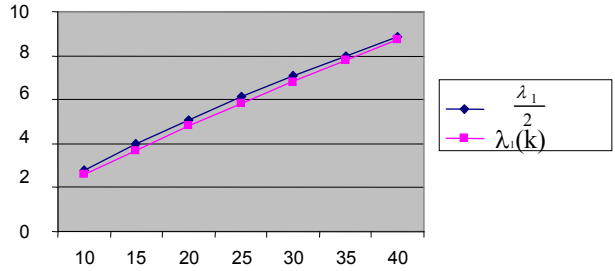

Fig. 5: Average delay of NRT packets versus R

## V. CONCLUSION

In this paper, we study an access control mechanism for RT and NRT packets arriving in a buffer implemented at an end user in HSDPA. The mechanism uses three thresholds to mange access in the buffer and gives access priority to RT packets. In order to control the arrival rate of the NRT packets in the buffer an active queue management is used.

In opposition of the reduction adopted in previous works where the NRT arrival rate is reduced to an arbitrary value, we propose a reduction which is a function of the total number of the packets in the buffer. And we show that this improve the performance parameters of NRT packets.

Mathematical description and analytical results are given, and numerical results show that:

- The mechanism with a linear reduction achieves lower delay for NRT in comparison with the model where the arrival rate of NRT packets is reduced to the mean value of a linear reduction, and this remain exact for every value of arrival rate of NRT packets before reduction, and every length of the space reserved to this type of packets;
- When we reduce the arrival rate of NRT packets to quarter, the mechanism with linear reduction isn't better only if the arrival rate of NRT packets before reduction is higher (case where congestion is more possible), and this still exact for every length of the space reserved to this type of packets.

## APPENDIX

$(\lambda_1 + \lambda)p(0,0) = \mu p(0,1) + \mu_1 p(1,0)$

$(\lambda + \lambda_1(a*L+b) + \mu_1)p(0,L) = \mu p(1,L) + \lambda_1 p(0,L-1) + \mu_1 p(0,L+1)$

$(\lambda + \mu_1)p(0,H) = \lambda_1(a*(H-1)+b)p(0,H-1) + \mu p(1,H)$

$(\lambda_1(a*L+b) + \mu)p(R,L-R) = \lambda p(R-1,L-R) + \lambda_1 p(R,L-R-1)$

$\mu p(R,H-R) = \lambda p(R-1,H-R) + \lambda_1(a*(H-1)+b)p(R,H-R-1)$

$(\mu + \lambda_1)p(R,0) = \lambda p(R-1,0)$

for $j=1,\ldots,L-1$: $(\lambda + \lambda_1 + \mu_1)p(0,j) = \mu p(1,j) + \lambda_1 p(0,j-1) + \mu_1 p(0,j+1)$

for $j=L+1,\ldots,H-1$: $(\lambda + \lambda_1(a*j+b) + \mu_1)p(0,j) = \mu p(1,j) + \lambda_1(a*(j-1)+b)p(0,j-1) + \mu_1 p(0,j+1)$

for $j=1,\ldots,L-R-1$: $(\lambda_1 + \mu)p(R,j) = \lambda p(R-1,j) + \lambda_1 p(R,j-1)$

for $j=L-R+1,\ldots,H-R-1$: $(\lambda_1(a*(R+j)+b) + \mu)p(R,j) = \lambda p(R-1,j) + \lambda_1(a*(R+j-1)+b)p(R,j-1)$

for $j=H-R+1,\ldots,H$: $\mu p(R,j) = \lambda p(R-1,j)$,

for $i=1,\ldots R-1$:

$(\lambda_1 + \mu + \lambda)p(i,0) = \lambda p(i-1,0) + \mu p(i+1,0)$

$(\lambda_1(a*L+b) + \mu + \lambda)p(i,L-i) = \lambda p(i-1,L-i) + \lambda_1 p(i,L-i-1) + \mu p(i+1,L-i)$

$(\mu + \lambda)p(i,H-i) = \lambda_1(a*(H-1)+b)p(i,H-i-1) + \mu p(i+1,H-i) + \lambda p(i-1,H-i)$

for $j=1,\ldots,L-i-1$: $(\lambda_1 + \mu + \lambda)p(i,j) = \lambda p(i-1,j) + \lambda_1 p(i,j-1) + \mu p(i+1,j)$

for $j=L-i+1,\ldots,H-i-1$: $(\lambda_1(a*(i+j)+b) + \mu + \lambda)p(i,j) = \lambda p(i-1,j) + \lambda_1(a*(i+j-1)+b)p(i,j-1) + \mu p(i+1,j)$

for $j=H-i+1,\ldots,H$: $(\mu + \lambda)p(i,j) = \lambda p(i-1,j) + \mu p(i+1,j)$

*The normalizing equation:* $\sum_{(i,j)\in E} p(i,j) = 1$